\documentstyle[12pt,aasms4,psfig]{article}

\begin{document}
\title{\bf New limits on  $\Omega_{\Lambda}$ and $\Omega_M$ from old galaxies 
at 
high redshift}
\author{J.S.Alcaniz\altaffilmark{1} and J.A.S.Lima\altaffilmark{2}}
\affil{Departamento de F\'{\i}sica, Universidade Federal do Rio Grande do 
Norte, 
\\ 
C.P. 1641, 59072-970, Natal, Brazil}
\altaffiltext{1}{alcaniz@dfte.ufrn.br}
\altaffiltext{2}{limajas@dfte.ufrn.br}
\begin{abstract}
The ages of two old galaxies (53W091, 53W069) at high redshifts are used to 
constrain the value of the cosmological constant in a flat 
universe ($\Lambda$CDM) and the density parameter $\Omega_M$ in 
Friedmann-Robertson-Walker (FRW) models with no $\Lambda$-term. In the case 
of 
$\Lambda$CDM models, the quoted galaxies yield two lower limits for the 
vacuum 
energy density parameter, $\Omega_\Lambda \geq 0.42$ and $\Omega_\Lambda \geq 
0.5$, respectively.  Although compatible with the limits from statistics of 
gravitational lensing (SGL) and cosmic microwave background (CMB), these 
lower 
bounds are more stringent than the ones recently determined using SNe Ia as 
standard candles. For matter dominated universes ($\Omega_\Lambda=0$), the 
existence of these galaxies imply that the universe is open with the matter 
density parameter constrained by $\Omega_M \leq 0.45$ and $\Omega_M \leq 
0.37$, 
respectively. In particular, these results disagree completely with the 
analysis 
of field  galaxies which gives a lower limit $\Omega_M \geq 0.40$. 
\end{abstract}
\pagebreak

In the last few years, the positive evidences against the standard FRW models 
are accumulating in a slow but increasing rate. Recent observations from a
large sample of type Ia supernovae are in contradiction with a universe 
closed
by ordinary matter ($\Omega_M \geq 1$), thereby appearing ruling out with
great confidence closed and flat FRW matter dominated universes. Indeed, even
open models, or more generally, any model  with positive deacceleration
parameter seems to be in disagreement with these data (Riess et al. 1998). In
the near future, one may expect that continued observations based on type Ia
supernovae as standard candles may determine the magnitude of the 
deceleration
parameter with great accurancy, as well as the value of the vacuum energy
density.

Another important piece of data is provided by the conflict between the 
expanding age of the universe, as inferred from measurements of the Hubble
parameter, and the age of the oldest stars in globular clusters. Recent
measurements of the Hubble parameter from a variety of techniques are now
converging into the range (one standard deviation) $h =
(H_{o}/100\rm{km s^{-1} Mpc^{-1}}) = 0.7 \pm 0.1$ (Freedmann et al. 1997, 
Mould et al. 1997, Nevalainen and Roos 1997). In particular, the current 
weighted value from the HST Key Project is $H_o=73 \pm 6$ (statistical)$ +      
8 $(systematic) $\rm{km s^{-1} Mpc^{-1}}$ (Friedmann 1998). This means that 
the
expansion age for a FRW flat matter dominated universe ($t_o = {2 \over
3}H_o^{-1}$) falls within the interval  $8.1 \rm{Gyr} \leq  t_o \leq 10.8
\rm{Gyr}$, while the ages inferred from globular clusters ($t_{gc}$) lie
typically in the range $t_{gc} \sim 13-15 \rm{Gyr}$, or even higher (Bolton
and Hogan 1995; Pont et al. 1998). Since the age of the universe for closed
models is even smaller than in the flat case, though some recent
determinations of $t_{gc}$ (based on the Hipparcos distance scale) have
decreased this value by approximately 2 Gyr (Chaboyer et al. 1998), the 
unique
possible conclusion is that the ``age crisis" continues  for closed, and at
least moderately, for flat FRW models. In this connection assuming a
reasonable incubation time for globular clusters ($\sim 1$ Gyr), Pont et al.
(1998) concluded that the minimum age of the Universe is 14 Gyr, thereby
ruling out a flat Universe, unless $h < 0.48$. All these results are also in
line with the latest constraints from SNe Ia, which provides $H_{o} = 65 \pm 
7
\rm{km s^{-1} Mpc^{-1}}$ with the uncertainty dominated by the systematic
errors commonly present in the calibration of the SN Ia absolute magnitude
(Riess et al. 1998). Actually, from the original matter dominated FRW class
with no cosmological constant, only extremely open universes may be old 
enough
to solve (beyond doubt) the expanding age problem. 

The already classical ``age problem" becomes even more acute if we consider 
the measurements of the age of the universe at high redshifts. For instance,
the $3.5$\rm{Gyr} radio galaxy (53W091) at $z=1.55$ discovered by Dunlop et
al. (1996) and confirmed by Spinrad et al. (1997) has been proved to be
incompatible with the age estimate for a FRW flat universe with no
cosmological constant unless the Hubble parameter is smaller than $45
\rm{km s^{-1} Mpc^{-1}}$ (Dunlop et al. 1996, Krauss 1997). Such a constraint 
is 
more
restrictive than  globular cluster age constraints.  Therefore, with the
exception of a very low Hubble constant variant, the  conclusion that the
standard FRW models seems to be inconsistent or only marginally compatible
with the estimated age of the universe is inescapable. 

On the other hand, a wide range of independent observations and 
theoretical arguments, closely related to inflationary scenarios, suggest 
that the more realistic model, which accommodates all the tests available at
present, is a flat universe with cosmological constant (Krauss and Turner,
1995). A positive cosmological constant significant today helps to solve the
``age problem" because it allows a period of cosmic acceleration at low
redshifts, thereby leading to expanding ages greater than the ones computed
with matter dominated universes. In particular, if $\Omega_\Lambda = 0.8$ and
$\Omega_M = 0.2$, the range of the age for the above considered values of $h$
is  $13.2\rm{Gyr} \leq t_o \leq 17.5 \rm{Gyr}$. However, although having
several independent positive evidences, the possibility of a non-zero
cosmological constant has not been proved beyond doubt and remains basically
an open question. In such a state of affairs, it is interesting to obtain
lower limits on the value of the vacuum energy density using different
methods. On the other hand, extremaly open models (OCDM) are at present the
main competitor of the $\Lambda$CDM models (Krauss 1998). Since the age of 
the
universe diminishes when $\Omega_M$ increases, the existence of old high
redshift galaxies will provide an upper limit to the density parameter. 

In the present work we discuss the constraints on $\Omega_\Lambda$ and 
$\Omega_M$ due to the existence of old high redshift galaxies (OHRG). Our 
study  differs from Krauss' work (1997) in two main aspects (see also Roos
and Harun-or-Rashid 1998). First of all,  we take into account a second  OHRG
recently reported (Dunlop et al. 1998). The second aspect is concerned with a
more methodological reason. Instead of analysing the constraints imposed by
the ages of these galaxies on the parameter space ($h, \Omega_M$) as done by
Krauss, we focus our analysis directly on the diagram $t(z)$ as a function of
$\Omega_\Lambda$. Our goal is to show how the estimated lower bounds on the
dimensionless``age parameter" of these two galaxies may be translated as 
lower
limits on the cosmological constant itself. As argued above, this approach 
may
also be applied to set upper limits for $\Omega_M$ in open universes with no
cosmological constant.      

The general age-redshift relation for FRW type
universes with  cosmological constant as a function of the observable
parameters is  \begin{equation}\label{agez} t(z) =
H_{o}^{-1}\int_{o}^{(1+z)^{-1}}\frac{dx}{\sqrt{1 - \Omega_{M}  + 
\Omega_{M}x^{-1} + \Omega_{\Lambda}(x^{2} - 1)}} \equiv
H_{o}^{-1}{f(\Omega_{M},  \Omega_{\Lambda}, z)} \quad, \end{equation} where
$\Omega_{M}$, $\Omega_{\Lambda}$, are the present day matter and  
cosmological
constant density parameters, respectively. If  $\Omega_{\Lambda} = 0$, the
above expression reproduces the well known result  for the standard model 
regardless of the value of $\Omega_M$ (Kolb and Turner 1990). For a flat
universe, equation ($\ref{agez}$) reduces to \begin{equation}\label{flatz}
t(z) = H_{o}^{-1} \int_{o}^{(1+z)^{-1}}\frac{dx}{\sqrt{(1 - 
\Omega_{\Lambda})x^{-1}  + \Omega_{\Lambda}x^{2}}} \quad, \end{equation} 
where
the matter density parameter $\Omega_M$ has been replaced in terms of 
$\Omega_{\Lambda}$ using the flat condition constraint $\Omega_M + 
\Omega_\Lambda =  1$. The above equation can be readily integrated (Gradstein
and Ryzhik, 1980)  yielding  
\begin{equation}\label{ageflat} t(z) =
\frac{2H_{o}^{-1}}{3\sqrt{\Omega_{\Lambda}}} \rm{ln} 
\left[\sqrt{\frac{\Omega_{\Lambda}}{1-\Omega_{\Lambda}}(1+z)^{-3}} + 
\sqrt{\frac{\Omega_{\Lambda}}{1 - \Omega_{\Lambda}}(1+z)^{-3} + 
1}\right]  \quad, \end{equation}
which should be compared with equation (2) in the paper by Krauss (1997). As 
one 
may check, in the limit $\Omega_{\Lambda}\rightarrow 0$ the age-redshift 
relation of the standard flat universe is recovered, namely
\begin{equation}\label{FFlat}
t(z) = \frac{2H_{o}^{-1}}{3(1 + z)^{3 \over 2}} \quad.
\end{equation}

Before discussing the resulting diagrams is useful to understand the method 
employed here. First of all, we take for granted that the age of the universe 
in 
a given redshift is bigger than or at least equal to the age of their oldest 
objects. As one may conclude, in the case of models with cosmological 
constant, 
the comparison of these two quantities implies in a lower bound for 
$\Omega_{\Lambda}$ since the age of the universe increases with the growth of 
the vacuum energy density. For standard matter dominated FRW models the same 
analysis  points to the opposite direction. In fact, the age of the universe 
increases when  $\Omega_{M}$ decreases, thereby implying the existence of an 
upper bound. In order to quantify these qualitative arguments, it proves 
convenient to introduce the ratio
\begin{equation}
\frac{t_{z}}{t_{g}} = \frac{f(\Omega_{M}, \Omega_{\Lambda}, z)}{H_o t_{g}} 
\geq 
1 
\quad ,
\end{equation}
where $t_g$ is the age of an arbitrary old object, say, a galaxy at the 
redshift 
$z$, and $f(\Omega_{M}, \Omega_{\Lambda}, z)$ is the dimensionless factor 
defined by (1). 

Notice that for each galaxy, the denominator of the above equation defines a 
dimensionless age parameter  $T_G=H_ot_g$. For the galaxy discovered by 
Dunlop 
et al. (1996), the lower limit to the age of this galaxy yields $T_{G}(1.55) 
= 
3.5H_o\rm{Gyr}$, which take values on the interval  $0.21\leq T_G \leq 0.28$. 
The extreme values of $T_G$ have been determined by the error bar of $h$. It 
thus 
follows that $T_G \geq 0.21$, and from (4) we see that at this $z$ the matter 
dominated flat FRW model furnishes an age parameter $T_{FRW} \leq 0.16$, 
which 
is far less than the previous value of $T_G$.  Naturally, for a given value 
of 
$h$, only models having an expanding age parameter bigger than the 
corresponding 
value of $T_G$ at $z=1.55$ will be compatible with the existence of this 
galaxy. 
In particular, the standard Einstein-de Sitter FRW model is ruled out by this 
test. 

In principle, the confidence on the limits derived here is ensured because we 
always consider the under estimation age of the galaxies (Peacock et al. 
1998) 
as well as the minimal value of ``little" h. We argue here that these two 
conditions for $t_g$ and $h$ give robust lower and upper limits for 
$\Omega_{\Lambda}$ and $\Omega_M$, respectively. The former condition is 
necessary because galaxies with small ages are more easily accomodated by the 
models, and the later because the age of the universe is inversely 
proportional to $h$, 
thereby also favouring the 
model. Naturally, similar considerations may also be applied to the $4.0$ Gyr 
old galaxy at $z=1.43$. In this case we have $T_G \geq 0.24$ while the matter 
dominated flat FRW model yields $T_{FRW} \leq 0.17$.

In Fig. 1 we show the diagrams of the dimensionless age parameter $T(z)= 
H_ot(z)$ as a function of the redshift for several values of 
$\Omega_{\Lambda}$. 
The forbidden regions in the graphs have been determined by taking the values 
of 
$T_G$, for each galaxy, separately. We see that the minimum values of 
$\Omega_\Lambda$ are, $\Omega_\Lambda \geq 0.42$ and 
$\Omega_\Lambda \geq 0.5$ providing a minimal total age of $12.9 \rm{Gyr}$ 
and
$13.5 \rm{Gyr}$, respectively. In Fig. 2, we present similar plots for matter
dominated universes. As should be physically expected, instead of lower 
bounds
as in the case of vacuum energy density, we now have upper limits to the 
value
of $\Omega_M$. The two galaxies lead to $\Omega_M \leq 0.45$ and $\Omega_M
\leq 0.37$ and provide a minimal total age of $12.4 \rm{Gyr}$ and $12.8
\rm{Gyr}$, respectively. As widely known, $\Lambda$CDM models are much more
efficient in solving  the ``age problem".

At this point, it is interesting to compare the results summarized in Table 1 
with some 
determinations of $\Omega_\Lambda$ and $\Omega_M$ 
 derived from
independent methods. As widely known, the current statistics of strong 
gravitational lenses 
provides powerful constraints on $\Lambda$CDM 
models. Recent studies predicted an upper limit $\Omega_\Lambda \leq 0.66$ at 
95.4$\%$ 
(2$\sigma$) confidence level (Kochanek 1996). This 
result, however, is not free of drawbacks, specially due to the presence of 
systematic 
errors, like extinction and the specific galaxy 
lens model. Recently, taking into account extinction, Falco, Kochanek and 
Mu\~{n}oz (1998) 
estimated that $\Omega_{\Lambda} \leq 0.7$ 
(2$\sigma$). More recently, this limit has again been slightly softened to 
$\Omega_\Lambda 
\leq 0.76$, at the same confidence level (Waga 
and Miceli 1998). If extinction is absent, the corresponding limit is 
$\Omega_\Lambda \leq 
0.55$ (2$\sigma$ level), which is slightly 
more stringent than the earlier Kochanek's bound. Therefore, by combining all 
these results 
with the OHRG lower bounds derived here, we 
find that the vaccuum density parameter probably lies in a very short range,  
$0.42 \leq 
\Omega_\Lambda \leq 0.7$. Assuming that the 
errors from extinction and galaxy lens model may be neglected or cancel out 
each other, the 
limit is more stringent, $0.42 \leq 
\Omega_\Lambda \leq 0.66$.  If the results of Waga and Miceli are considered, 
the 
corresponding ranges, with and with no 
extinction, are $0.42 \leq \Omega_\Lambda \leq 0.76$ and $0.42 \leq 
\Omega_\Lambda \leq 
0.55$ (2$ \sigma$ level), respectively. Notice 
that our lower bound is less severe than $\Omega_\Lambda \geq 0.60$ obtained 
by Roos and 
Harun-or-Raschid (1998).

As stated before, a more satisfactory evidence for a positive 
$\Omega_\Lambda$ as well as 
for an accelerated state of the present 
universe, has been obtained from distant SNe Ia (Riess et al. 1998). The 
combination of data 
from a large sample of 48 SNe Ia point 
(nearby and at high $z$) favors consistently  an expanding universe dominated 
by a positive 
cosmological constant, e.g., a negative 
deceleration parameter 
($q_o = {\Omega_M \over 2} -\Omega_\Lambda < 0$). From  this set of data the 
values of the 
density parameters are $\Omega_\Lambda =  
0.72^{+0.72}_{-0.48}$ and $\Omega_M = 0.24^{+0.56}_{-0.24}$ (see Table 1). 
The main 
conclusion is that a non-negligible vacuum energy 
density is extremely probable. In a recent paper, Perlmutter et al. (1998) 
also obtained 
$\Omega_\Lambda = 0.4 \pm 0.2$, that is, 
$\Omega_\Lambda \geq 0.2$. (see Table 1). Note that both lower bounds, 
namely, 
$\Omega_\Lambda \geq 0.24$ and $\Omega_\Lambda \geq 0.2$ 
are much less stringent than $\Omega_\Lambda \geq 0.42$ derived here. There 
is, however, a 
possiblity that these two lower bounds for 
$\Omega_\Lambda$ (from OHRG and SNe Ia) might become compatible in the near 
future. For 
example, suppose that due to some physical 
effect, which has been unnoticed so far or simply not taken into account in 
the estimates, 
the lower limits to the age of these objects 
are smaller than 3.0 Gyr. This means that the lower limit of $\Omega_\Lambda$ 
is 
proportionaly smaller than the ones derived here. Let us 
now assume a more conservative lower bound than the ones provided by SNe Ia 
observations, 
say, $\Omega_\Lambda \geq 0.14$. In this case, 
using the analytical formula (3), one may check that the age of the galaxy 
53W091 should be 
$2.6$ Gyr. However, if the lower limit of 
$\Omega_\Lambda$ is $0.2$ as claimed by Perlmutter et al. (1998) or $0.24$ 
(Riess et al. 
1998), the ages of this galaxy are $2.8$ Gyr and 
$2.86$ Gyr, respectively. Parenthetically, with a high level of confidence, 
the age of this 
object is bigger than $2.5$Gyr (Dunlop 1999, 
private communication). In the previous analysis, we have implicitly supposed 
that possible 
evolutionary effects of SNe Ia are completely 
under control, thereby giving rise to very robust constraints. Naturally, if 
the opposite 
viewpoint is assumed, the range allowed for 
$\Omega_{\Lambda}$ combining OHRG and SLG
data is maintained, and some basic aspect of the standard SNe Ia approach
deserves a closer scrutiny.    

In the case of a matter dominated universe ($\Omega_\Lambda = 0$), 
some authors have recently argued that an undercritical density with
$\Omega_M \sim 0.5$ should be adopted (Willick et al. 1997; Tammann 1998; da
Costa et al. 1998). These claims are also in contradiction with the upper
limits for $\Omega_M$ from old galaxies at high $z$ determined in the present
paper (see Table 1). In this connection, as pointed out recently, the
observational properties of 53W091 and the findings of galaxy candidates at 
$z\gg 5$ 
strongly suggest that the Universe contains more 
collapsed galaxies
than the modified CDM model would predict (Kaslinsky and Jimenez 1998).

These results reinforce the interest in searching for old galaxies 
at high redshifts. Special attention should be given for those galaxies where
the stellar population was born at once, that is, with no further starburst
formation. By definition these are the best clocks, and hopefully, this kind
of object may be identified in the near future. The overall tendency is that
if more and more galaxies are discovered, the relevant statistical studies in
connection with the ``age problem" may provide very restrictive constraints
for any  realistic cosmological model. Conversely, given a population of
galaxies at high $z$, the ages of which are not well known, the method
discussed here may readily be inverted yielding upper limits to the age of 
the
galaxies once a reasonable model of the universe is adopted. 

\acknowledgments
\section*{Acknowledgments}

The authors are grateful to R. Brandenberg, R. Opher, M. Roos and I. Waga   
for 
helpful discussions. This work was 
partially suported by the Conselho Nacional de Desenvolvimento 
Cient\'{\i}fico e 
Tecnol\'{o}gico (CNPq) and Pronex/FINEP (no. 41.96.0908.00). 


\newpage

\section*{Table 1}

\begin{table}[t]
\begin{center}
\begin{tabular}{rlll} \hline \hline
\multicolumn{1}{c}{Method}&
\multicolumn{1}{c}{Autor(s)}&
\multicolumn{1}{c}{$\Omega_{\rm{M}}$}& 
\multicolumn{1}{c}{$\Omega_{\Lambda} \rm{(flat)}$}\\
\hline
SNe Ia& Perlmutter et al. (1998)& $0.2 \pm 0.4$& $0.4 \pm 0.2$\\
SNe Ia& Riess et al. (1998)& $0.24^{+0.56}_{-0.24}$& 
$0.72^{+0.72}_{-0.48}$\\
FG& Tammann (1998)& $\sim 0.5$&  \quad     - \\
SGL& Kochanek (1996)& $> 0.15$& $< 0.66$\\
CMB & Lineweaver (1998) & $0.24 \pm 0.1$ & $0.62 \pm 0.16$ \\
OHRG ($z = 1.43$) & this paper& $\leq 0.37$ & $\geq 0.5$\\
OHRG ($z = 1.55$) & this paper& $\leq 0.45$ & $\geq 0.42$\\
\hline \hline
\end{tabular}
\caption{Limits to $\Omega_{\rm{M}}$ and $\Omega_{\Lambda}$} 
\end{center}
\end{table}

\pagebreak


\begin{figure}
\vspace{.2in}
\centerline{\psfig{figure=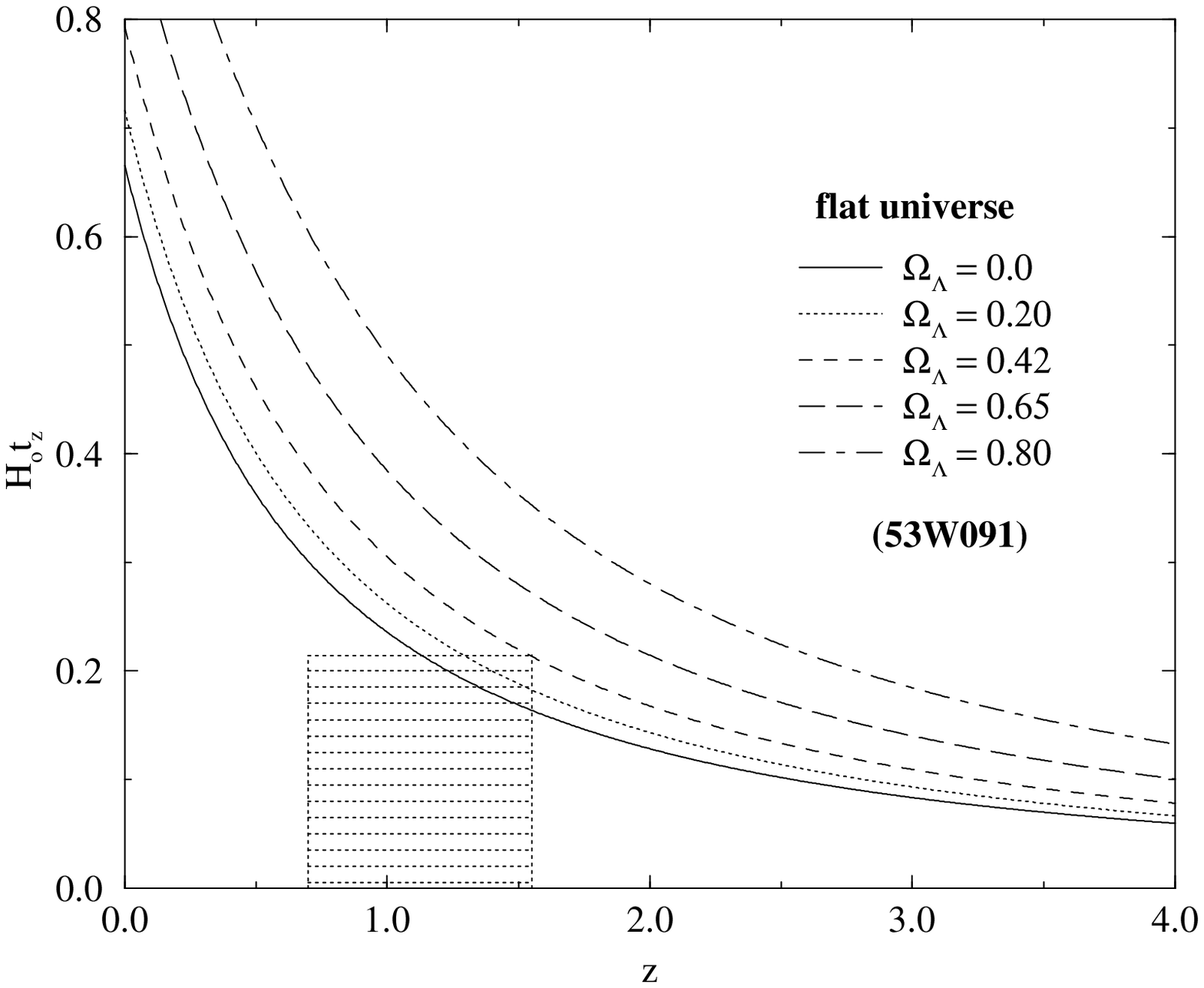,width=3.5truein,height=3.5truein}
\hskip 0.1in
\psfig{figure=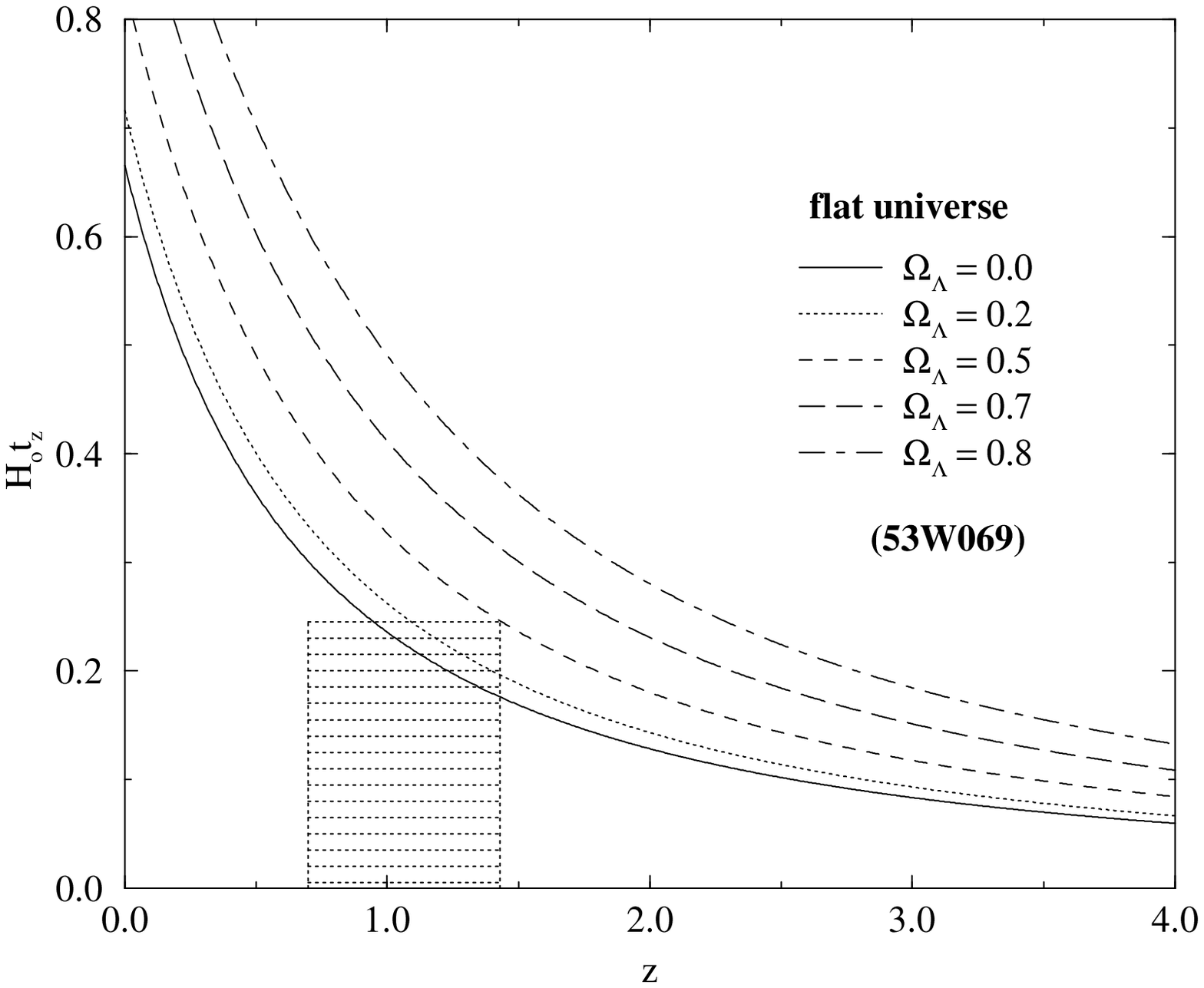,width=3.5truein,height=3.5truein}}
\caption{The dimensionless age parameter as a function of the redshift for some
selected values of $\Omega_{\Lambda}$.  All curves crossing the dark shading
rectangle yield an age parameter less than the minimal value required by the
galaxies discovered by Dunlop et al. (1996 and 1998).  As expected, this class includes the
FRW flat universe with $\Omega_{\Lambda}=0$ (solid curve).  The curve
intersecting the corner of the rectangle corresponds to $\Omega_{\Lambda}=0.42$ (53W091)
and $\Omega_{\Lambda}=0.5$ (53W069) and fix the minimal value of $\Omega_\Lambda$ allowed by these two galaxies.}
\label{fig1}
\end{figure}

\begin{figure}
\vspace{.2in}
\centerline{\psfig{figure=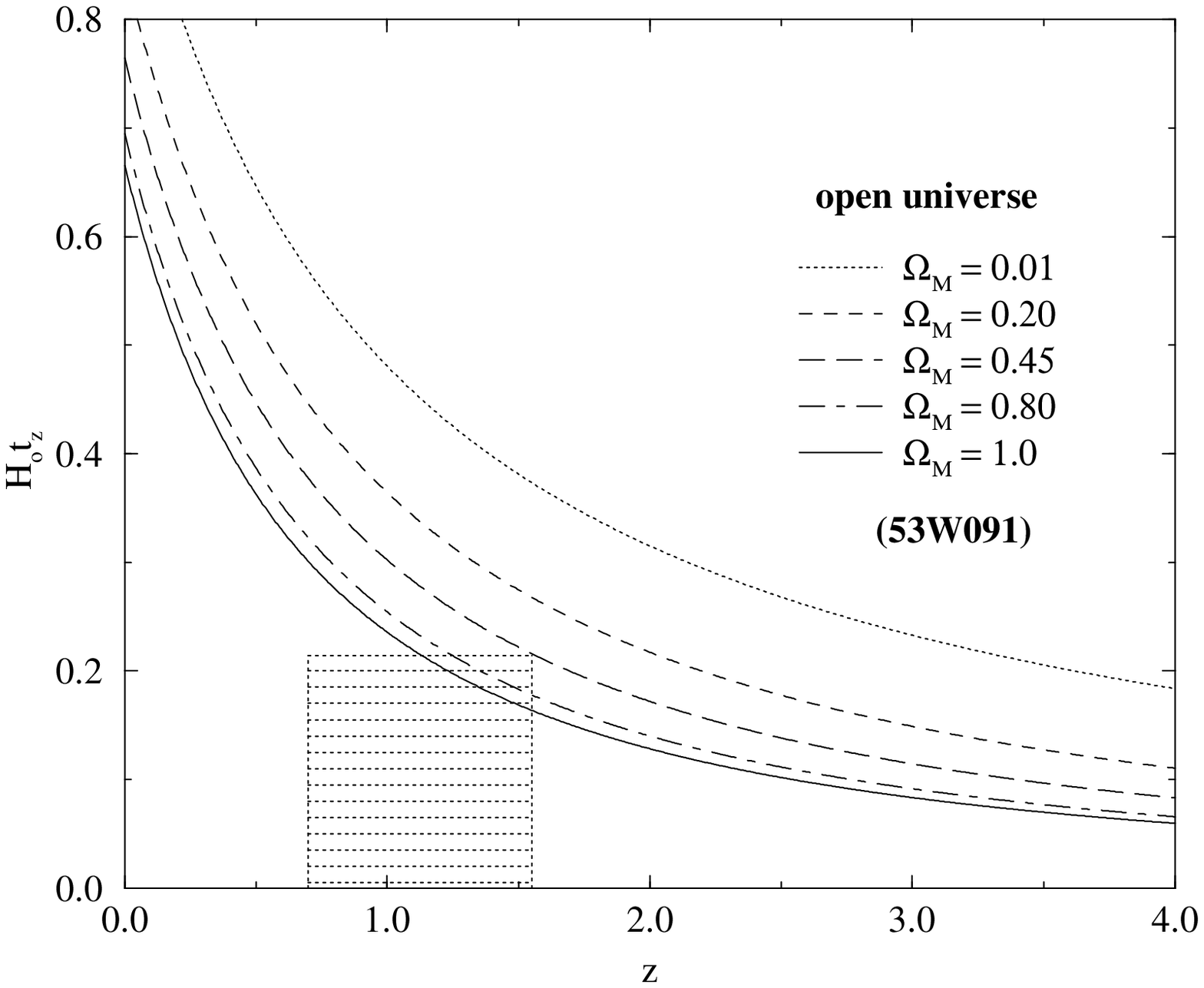,width=3.5truein,height=3.5truein}
\hskip 0.1in
\psfig{figure=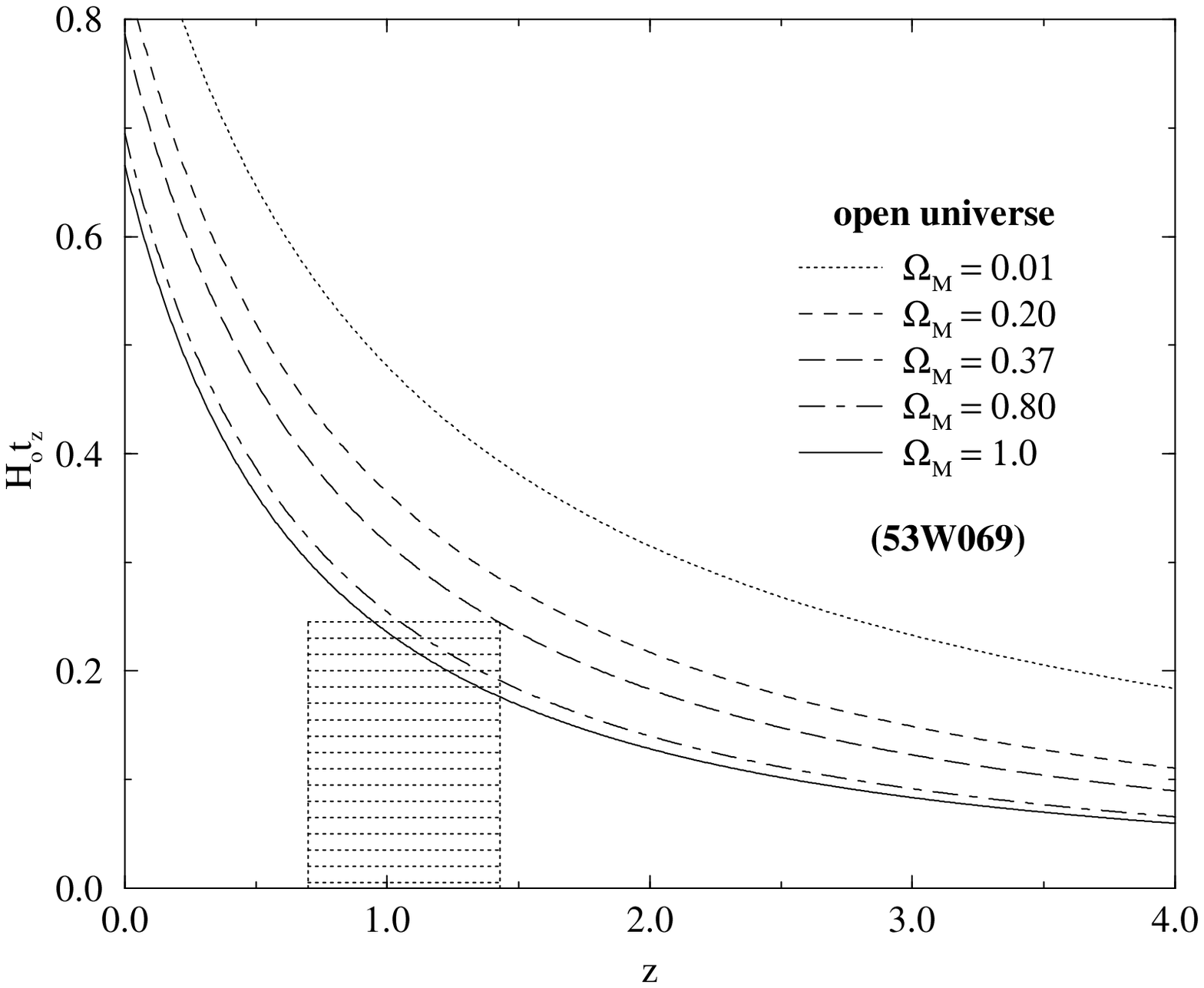,width=3.5truein,height=3.5truein}}
\caption{The dimensionless age parameter as a function of the redshift for FRW universes 
with no cosmological constant. As before, all curves crossing the dark shading rectangle 
yield an age parameter less than the minimal value required by the galaxy discovered by
Dunlop et al. (1996 and 1998). This subset includes the standard FRW flat universe (solid curve). 
Clearly, $\Omega_{o}=0.45$ is the maximum value of $\Omega_o$ allowed by 53W091 and $\Omega_{o}=0.37$ by 53W069.}
\label{fig2}
\end{figure}

\end{document}